\documentclass[twocolumn,aps,pra,superscriptaddress]{revtex4-1}
\usepackage{graphicx}
\usepackage{natbib}
\usepackage{url}
\usepackage{textcase}
\usepackage{bm}
\usepackage{color}
\usepackage{times}
\usepackage{siunitx}
\usepackage{ amssymb }

\begin{document}
%\title{CMOS architecture for 2D spin qubit array}
%\title{Large-scale quantum computation with silicon quantum dot qubits}
\title{Silicon CMOS architecture for a spin-based quantum computer}

\author{M. Veldhorst}
\affiliation{Qutech, TU Delft, 2600 GA Delft, The Netherlands}
\affiliation{Centre for Quantum Computation and Communication Technology, School of Electrical Engineering and Telecommunications, The University of New South Wales, Sydney, NSW 2052, Australia}
\author{H.G.J. Eenink}
\affiliation{Centre for Quantum Computation and Communication Technology, School of Electrical Engineering and Telecommunications, The University of New South Wales, Sydney, NSW 2052, Australia}
\affiliation{NanoElectronics Group, MESA+ Institute for Nanotechnology, University of Twente, P.O. Box 217, 7500 AE Enschede, The Netherlands}
\author{C.H. Yang}
\affiliation{Centre for Quantum Computation and Communication Technology, School of Electrical Engineering and Telecommunications, The University of New South Wales, Sydney, NSW 2052, Australia}
\author{A.S. Dzurak}
\affiliation{Centre for Quantum Computation and Communication Technology, School of Electrical Engineering and Telecommunications, The University of New South Wales, Sydney, NSW 2052, Australia}

\date{\today}
%\pacs{}
\maketitle

\textbf{
Recent advances in quantum error correction codes for fault-tolerant quantum computing and physical realizations of high-fidelity qubits in a broad range of platforms give promise for the construction of a quantum computer based on millions of interacting qubits. However, the classical-quantum interface remains a nascent field of exploration. Here, we propose an architecture for a silicon-based quantum computer processor based entirely on complementary metal-oxide-semiconductor (CMOS) technology, which is the basis for all modern processor chips. We show how a transistor-based control circuit together with charge-storage electrodes can be used to operate a dense and scalable two-dimensional qubit system. The qubits are defined by the spin states of a single electron confined in a quantum dot, coupled via exchange interactions, controlled using a microwave cavity, and measured via gate-based dispersive readout. By implementing a spin qubit surface code, we show that this silicon quantum processor architecture can enable universal quantum computation.}

\begin{figure*}
\includegraphics[width=\textwidth]{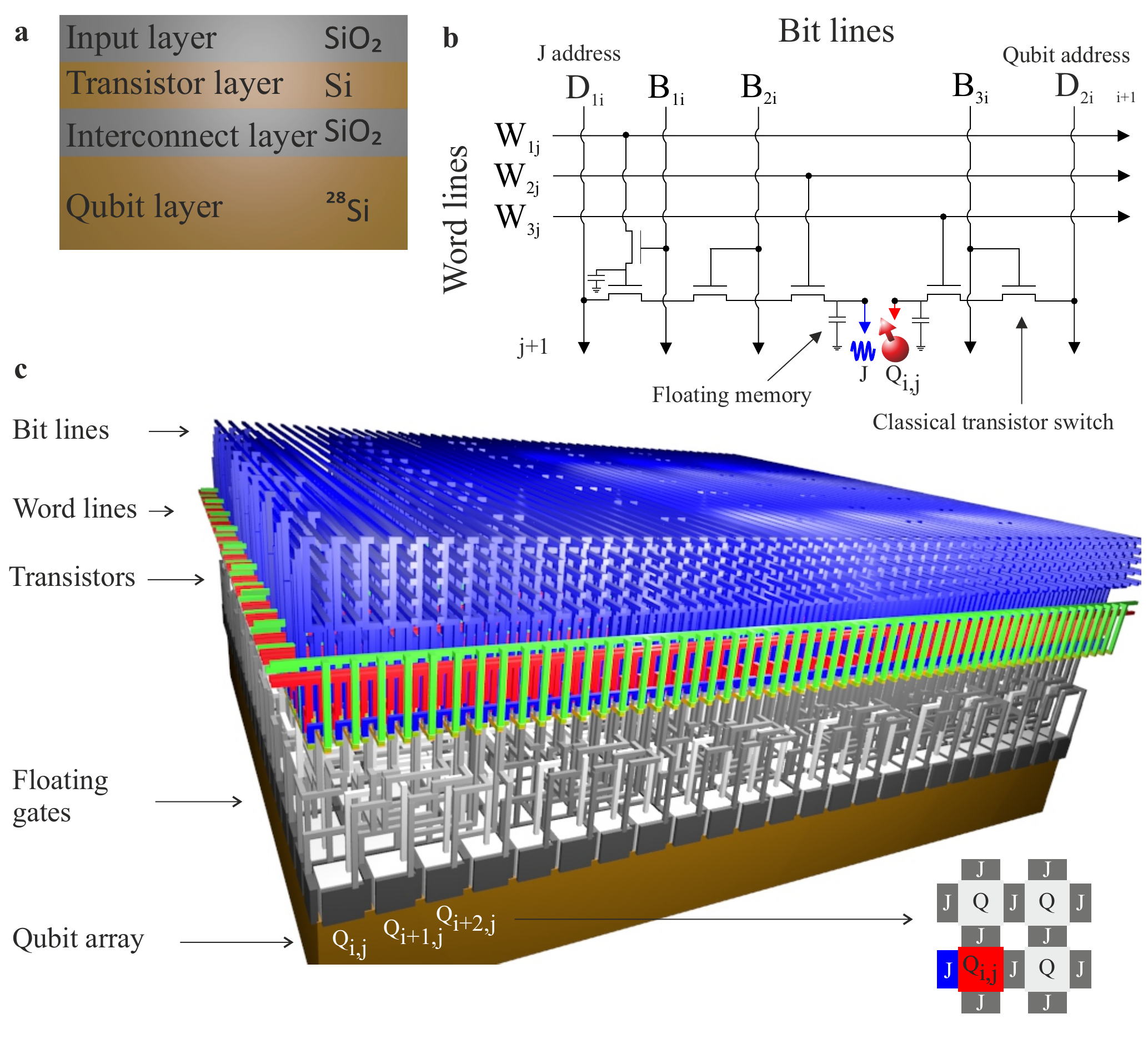}%
\caption{ \textbf{Physical quantum processor. a} A silicon-on-insulator (SOI) wafer is processed, such that the bottom layer of isotopically enriched silicon-28 contains the 2D qubit array and the top layer of silicon forms the transistors to operate the qubits. These are interconnected through the oxide regions using polysilicon vias. \textbf{b} Electrical circuit for the control of one $Q$-gate and one $J$-gate allowing the required individual, row-by-row, or global operations, as explained in the main text. \textbf{c} Physical architecture to operate one unit module containing 480 qubits. The inset on the bottom right shows a plan view cross-section through the qubit plane. Each $J$ gate and qubit is connected via the circuit shown in \textbf{(b)}. \label{fig:1}}
\end{figure*}

%\section{Introduction}
The most promising routes towards large-scale universal quantum computing all require quantum error correction (QEC) \cite{Terhal2015}, a technique that enables the simulation of ideal quantum computation using realistic noisy qubits, provided that the errors are below a fault-tolerant threshold. Using the most forgiving methods, such as the two dimensional surface code \cite{Dennis2002}, these error thresholds can be as high as 1\% \cite{Fowler2012}, a level that is now routinely achieved across several qubit platforms \cite{Kok2007, Brown2011, Barends2014, Waldherr2014, Dolde2014, Muhonen2014, Veldhorst2014}. However, these approaches also require a platform that can be scaled up to very large numbers of qubits, of order $10^8$. This currently creates one of the most stringent barriers in the field, even for the most promising platforms. Here, we propose a method to overcome this hurdle, using spin qubits in silicon and taking direct advantage of CMOS technology. While silicon was recognized early on as a promising platform in the seminal work of Kane \cite{Kane1998}, leading to many novel architectures \cite{Hollenberg2006, Trifunovic2012, OGorman2016, Hill2015, Tosi2015, Pica2016, Jones2016}, a key and contrasting feature of our approach is that each architectural component is based on existing devices and commercially available technology to provide a scalable solution. We show that it is possible to construct a highly dense two-dimensional qubit array starting from a single silicon-on-insulator (SOI) wafer. 

Silicon CMOS integrated circuits (ICs) are the prototypical example for scalable electronic platforms, now holding transistor counts exceeding billions. This remarkable level of integration is based upon decades of advances in silicon materials technologies, and these will also be crucial in the development of high-quality spin qubits. A key architectural aspect of ICs has been the use of parallel addressing via word lines and bit lines facilitating rapid read and write operations on large 2D arrays of bits. Unfortunately, this method cannot directly be applied to scale qubit arrays. Unlike transistors, the tolerance levels of qubits are small, thereby requiring individual tunability. However, as we show here, the highly repetitive nature of error correction methods like the surface code enables the use of an advanced protocol for parallel addressing. Individual qubit stabilization is further obtained via floating memory gate electrodes that can be routinely reset, similar to dynamic random access memory (DRAM) systems. Together, these allow the design of a platform where the number of addressing lines increases in a scalable manner proportional to $\sqrt{N}$, where $N$ is the number of qubits.

\section{Physical architecture}
The general architecture we propose is depicted in Fig. \ref{fig:1}. We start with a SOI wafer, where the top layers host the classical circuitry, the isotopically enriched silicon-28 bottom layer holds the quantum circuit, and these are interconnected via metal lines which penetrate the oxide region, see Fig. \ref{fig:1}a. The fabrication could be performed monolithically, from a single wafer, or include flip-chip technologies to enable the construction of the two circuits separately. We focus here on single spin qubits confined in quantum dots \cite{Veldhorst2014}. For complete qubit control, one data line ($D_{2i}$) is interconnected to each corresponding qubit ($Q_i$) to tune the qubit resonance frequency ($\nu_i$), while a second ($D_{1i}$) interconnects to each $J$-gate to control the exchange coupling between qubits, shown in Fig. \ref{fig:1}b. To provide individual, row, or global qubit addressing, the data lines are controlled by a combination of word lines ($W$) and bit lines ($B$). Assuming the minimal width of, and separation between, the gates and doped regions is equal to the minimum feature size $\lambda$, the classical circuit occupies an area 80$\lambda^2$ per qubit (see Supplementary Information section 1 for further details). A feature size of 7nm would require a minimum qubit size of $\approx$ 63 nm $\times$ 63 nm, consistent with experimental realizations of silicon quantum dot qubits \cite{Veldhorst2014, Kawakami2014}. Large foundries are now capable of manufacturing some features down to this size, but ongoing advances in down-scaling will be needed to fabricate the classical devices assumed here, and so the development of such a quantum computer will therefore need to proceed hand-in-hand with the ongoing advances in semiconductor technology.

Generally, the most compact classical circuits have different geometries from quantum circuits. The situation is further complicated by the geometrical layout of the metal connection lines, determined by the quantum error correction implementation. We have overcome the complexity in scaling these differently sized circuit components via the use of vertically-stacked interconnection layers, and as the number of qubits increases, the three layers become spatially identical. This point is reached upon expanding the structure to host 480 qubits, as shown in Fig. \ref{fig:1}c (see the Supplementary Information section 2 for further details), and beyond this further scaling becomes a straightforward replication of this 480 qubit module. A full quantum processor would then contain multiple modules and the edges would be connected to a doped silicon region, serving as an electron reservoir, from which electrons may be sequentially loaded into the qubit array as is done in charge-coupled devices \cite{Baart2016}. The word and bit lines of the integrated quantum processor chip will then be connected to classical control and measurement electronics \cite{Reilly2015} that can reside next to or further away from the quantum chip depending on their level of power dissipation. 

\begin{figure*}
\includegraphics[width=\textwidth]{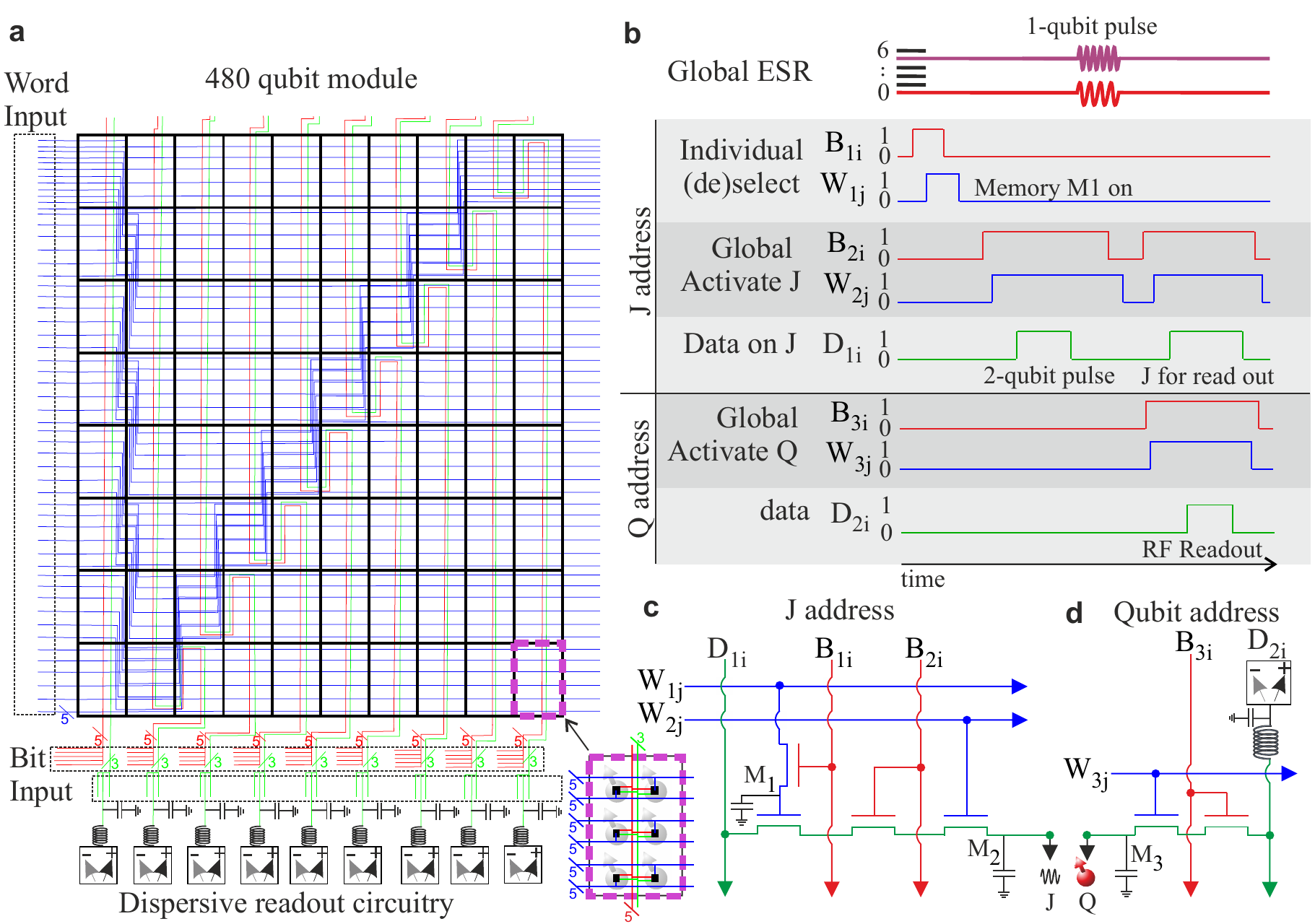}%
\caption{\textbf{Electrical circuit and qubit addressing scheme. a} Electrical wiring of the 480 qubit module. The word lines ($W$), bit lines ($B$) and data lines ($D$) can be addressed to enable global control, to couple and readout row-by-row and to individually (de)select qubits. The $W$ and $B$  lines are grouped in five and the $D$ in three, such that a combination of these form the lines of the electrical circuit of a single extendable structure, consisting of a single qubit and two $J$ gates. The zigzag structure in (\textbf{a}) is to accommodate for the different aspect ratios of qubit size, control size and in order to be consistent with surface code operation (see Supplementary Information section 2 and 3 for further details). The purple rectangle displays the region that is occupied by 6 qubits, corresponding to a surface code unit cell (see Fig.3a). Note that the word lines are connected to the qubits in an alternating arrangement in order to make the circuit compatible with our spin qubit surface code scheme. \textbf{b} Typical operation protocol of the electrical circuit shown in \textbf{c} and \textbf{d}. Individual qubit selection can be via lines $W_{1}$ and $B_{1}$ that (de)charge floating electrodes ($M1$ in \textbf{c}) and (dis)connect the data lines from the corresponding $J$-gates. Two-qubit operations are performed by activating the associated lines $W_2$ and $B_2$ and sending a pulse through data line $D_1$. Global single-qubit operations can be applied by broadcasting an ESR pulse at the resonance frequencies of the corresponding subgroup of qubits at any time of the sequence. Readout is enabled via the lines $W_2$, $B_2$, $W_3$, and $B_3$. Then a pulse turns on the selected $J$ gates, and RF readout is performed via the data line $D_2$ connected to the qubit. The electrical circuits in \textbf{c} and \textbf{d} show the corresponding structures to control the qubits and the exchange coupling between them. The floating memories $M_1$ and $M_2$ are to maintain the desired electric fields on the respective $J$ and $Q$ gates and may be periodically refreshed.}
\label{fig:2}
\end{figure*}

\section{Electrical operation}
\label{section:readout}
\begin{figure*}
\includegraphics[width=\textwidth]{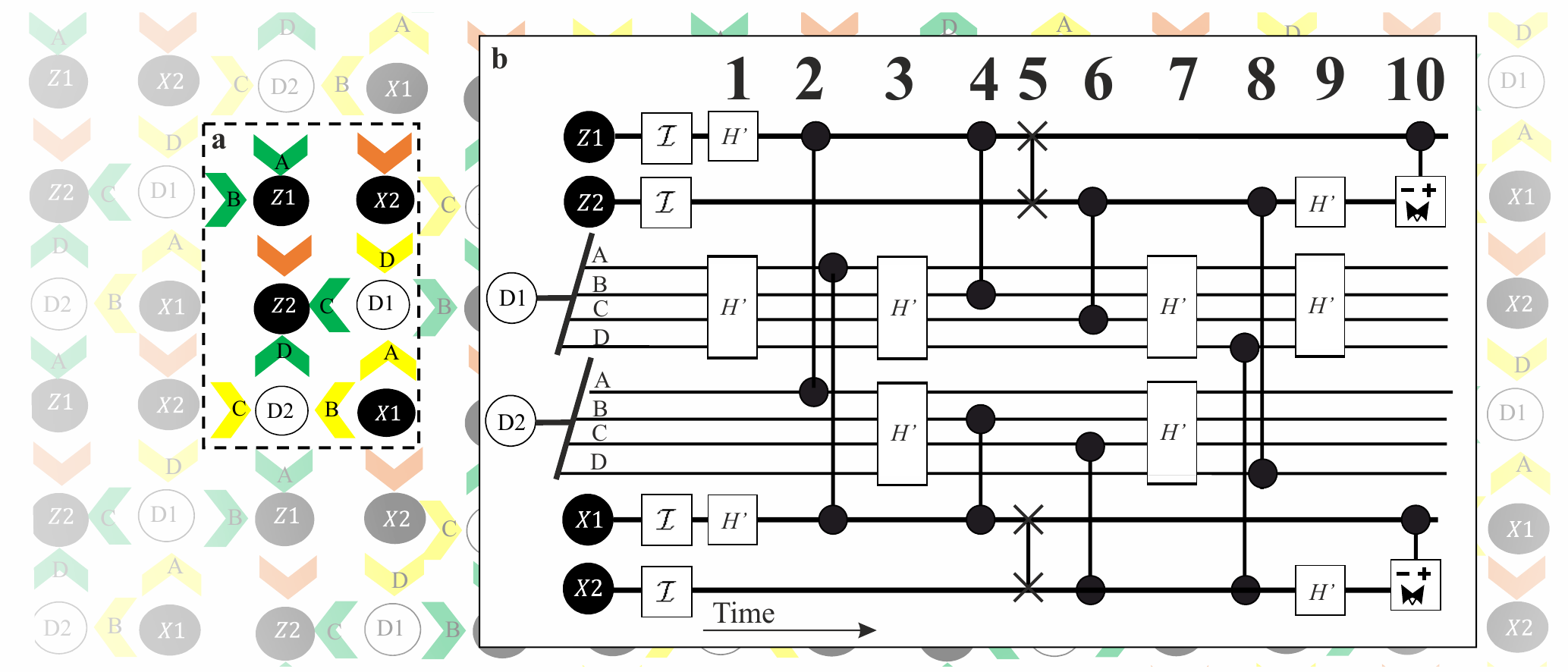}%
\caption{\textbf{Surface code operation. a} Single unit cell, containing six qubits: two data qubits, $D_1$ and $D_2$, and four measurement qubits, $Z_1$, $Z_2$, $X_1$ and $X_2$. Each of these qubit classes has a well-defined independent qubit resonance frequency. \textbf{b} Surface code operation based on this unit cell. Note that the labels $A,B,C$ and $D$ refer to the data qubits associated with the respective measurement qubit (see also \textbf{a}). A single cycle of initialization, control, and readout corresponds to ten steps. An additional SWAP operation is included, compared to standard surface code operation \cite{Fowler2012}, for the qubit readout step. The single qubit operations $H'$ are Hadamard-like and compensate for $\hat{z}$ axis rotations that can occur during the CPHASE \cite{Veldhorst2015}. See Supplementary Information section 3 for qubit operation and for details on comparison between general surface code and this quantum dot surface code. 
\label{fig:3}}
\end{figure*}

We now turn to the electrical operation of the qubit module, Fig. \ref{fig:2}, and consider surface code operation, Fig. \ref{fig:3}. A single electron is loaded into each quantum dot by addressing the corresponding word and bit lines. The electron occupancy is verified by gate-based dispersive readout, as shown in Fig. \ref{fig:2}c and described further below. We assume that the complete structure is maintained at cryogenic temperatures ($\sim$1 K or less) inside an electron spin resonance (ESR) cavity, which will be used to apply qubit control pulses. Each qubit must be calibrated to its desired qubit resonance frequency by tuning the associated floating memory gate, using electrical $g$-factor control, as has been demonstrated experimentally \cite{Veldhorst2014}. The surface code operation we discuss here requires a total of six different resonance frequencies (see Fig. \ref{fig:3}). The qubit gates ($Q_{ij}$) are calibrated to voltages such that the exchange coupling between adjacent qubits is negligible when the intermediate $J$-gates are set at an "off" bias point, and for which there is a common value of exchange when the $J$-gates are set to an "on" bias. Global (i.e. parallel) control is a crucial aspect for large-scale operation. The use of floating memory gates in the proposed architecture here has the significant advantage of enabling the individual tuning of qubits, while having a minimal amount of control lines that can then be set to common bias levels, thus enabling global operations. 

\subsection{Gate-based dispersive readout and initialization}
Two popular methods for spin qubit readout are based on spin to charge conversion: readout based on the Zeeman energy (using a reservoir) \cite{Elzerman2004} and readout based on the singlet-triplet energy (via Pauli spin blockade) \cite{Petta2005}. Both approaches can be made compatible with our control circuit, but readout based on Pauli spin blockade offers a number of advantages, including: a larger energy scale leading to higher readout fidelity; no necessity for a large electron reservoir; and a large magnetic field is not required so that the qubit resonance frequency can be freely chosen (e.g. operation can be at a rather low frequency of order one GHz). We therefore propose to use Pauli spin blockade for parity readout between two spin qubits. 

Dispersive readout \cite{Colless2013} is generally considered for multi-dot qubits such as singlet-triplet qubits \cite{Petta2005}, but here we envision the readout of single spins by exploiting Pauli spin blockade. Single spin states can be projected onto singlet-triplet states using a reference neighbour dot, thus allowing a parity measurement between two qubits. Starting from the (0,2) singlet ground state, qubit initialization is obtained by adiabatically moving to the (1,1) state, which results in a spin-down state in the dot with the larger $g$-factor \cite{Veldhorst2015}. The adiabaticity here is with respect to both the tunnel coupling and resonance energy difference between the qubits, which can be larger than 100MHz \cite{VeldhorstPRB2015}.  Qubit readout occurs via the reverse process of initialization. Depending on the target qubit spin state, the (0,2) singlet state will be partly occupied. This will result in a capacitance that is dependent on the state of the target qubit, which can be observed in the reflected power in the RF circuit connected to a nearby gate \cite{Colless2013}, see Fig. \ref{fig:2}.

The readout is performed in a row-by-row manner and the parity analyzers are connected to the data lines $D_{2i}$ via bias tees, see Fig. \ref{fig:2}d. Using classical circuitry, it is possible to frequency multiplex an entire row \cite{Hornibrook2014} so that only one RF analyzer circuit is needed, however it could be more convenient to use separate analyzers for each bit line, as depicted in Fig. \ref{fig:2}a. Operating dispersive readout at 1 GHz enables readout on the order of 10-100 ns, such that a large qubit array can be read out well within the single qubit coherence time of 28ms in $^{28}$Si substrates \cite{Veldhorst2014}. 

To be able to perform parallel operations, an integrated 3D arrangement of the addressing and qubit structures is required, such that a certain combination of word lines and bit lines will address the same particular qubit in each unit cell. This is implemented in the schematic in Fig.\ref{fig:2}, where the unit cell has a size 2x3 (2 data qubits and 4 measurement qubits). Then, 20x24 qubits may be addressed using input lines on a grid 9x54 (see Supplementary Information section 2). To deselect individual qubits, the $J$-gates surrounding the relevant qubits are deactivated, thereby isolating them from the data qubits and creating an additional degree of freedom in the array for quantum computation. This protocol will be particularly relevant for operation of the defect-based surface code, as described in section \ref{section:readout}.b.

\subsection{Surface code operations}
Surface codes are among the most promising methods for quantum error correction \cite{Terhal2015, Fowler2012}. In our approach, an alternating arrangement of data and measurement qubits is used, where two data qubits interact with four measurement qubit neighbours, and the surface code unit cell becomes as shown in Fig.\ref{fig:3}a. Two measurement qubits together enable a parity readout step, and this implementation is thus slightly larger than the usual surface code unit cell of four qubits. The measurement qubits are initialized to $\mathcal{I}$ by adiabatically moving from the (0,2) charge state to the (1,1) charge state, as discussed in section \ref{section:readout}.a. Single qubit operations and the two-qubit CPHASE and SWAP operations are then performed, followed by dispersive readout. The complete surface code cycle for quantum dot qubits, see Fig.\ref{fig:3}b, then involves ten steps.

The focus of the work presented here is the realization of a 2D qubit array and we envision that many surface code schemes and even analog quantum simulator algorithms can be constructed based on our design. We therefore do not undertake here a detailed analysis of the particular error thresholds associated with our surface code implementation. However, since our implementation is based on general surface codes and the number of operations is comparable with those previously reported \cite{Fowler2012}, we expect that the fault-tolerant error thresholds will largely remain the same; see Supplementary Information section 3 for a comparison between general surface codes and the spin qubit surface code as presented here. Recent demonstrations of single- and two-qubit gates in silicon \cite{Veldhorst2014, Veldhorst2015} provide thereby significant scope to meet all the required fault-tolerant thresholds. Further improvements in two-qubit fidelities fidelities are conceivable, for example via operation at the charge symmetry point for a pair of quantum dot qubits \cite{Reed2016, Martins2016}.

To perform logical quantum operations on the qubit module with a defect-based surface code, qubit deselection is required to create holes for braiding operations \cite{Fowler2012}. Individual qubit (de)selection is enabled by the circuit shown in Fig. \ref{fig:2}c, using word and bit lines $W_{1j}$ and $B_{1i}$. The required holes will be limited, as most physical qubits will be used to create the logical qubits. The infrequent nature of required qubit (de)selection allows for this to be done individually, rather than globally, and we achieve this by deactivating the associated $J$-gates, thereby isolating the associated data qubits from their measurement qubits.

\section{Heat dissipation}
A critical factor for almost any large-scale computing platform is cooling power. While it is not within the scope of this manuscript to calculate the total power dissipation that will depend on the exact layout of the architecture, we estimate here the dynamic power produced by the $J$-gates, which is likely the largest source of dissipation. The power dissipation of a single surface code unit cell, shown in Fig. \ref{fig:3}, is given by $P= C V^2 \alpha f$,  with $C$ the capacitance for the floating memory, $V$ the switching voltage, and $\alpha$ the activity factor relative to the surface code clock cycle with frequency $f$ $\approx$ \SI{0.1}{\mega \hertz} (assuming Rabi frequencies on the order of 1MHz \cite{Veldhorst2014}). The surface code unit cell is operated using 54 transistors and during a full cycle the $J$-gate actvity $\alpha$ = 12. The floating gate electrodes may be periodically refreshed, as in DRAM technology, but we estimate that for high-fidelity qubit operation $RC$ times beyond one second will be required to avoid significant drifts during operation. We assume this requires a capacitance $C$ $\approx$ \SI{1}{\pico \farad}, with an associated Johnson-Nyquist thermal noise $V_{thermal}=\sqrt{K_BT/C}\approx$ \SI{1}{\micro\volt}, providing a tolerable level \cite{Veldhorst2015}. Assuming a switching voltage $V$ = 0.2 V results then in a power dissipation for a single unit cell of $\approx$ \SI{50}{n\watt}.

The available cooling power depends on the dilution refrigerator, but will ultimately be limited by the thermal conductivity of the addressing lines in the upper layers of the circuit. The thickness will depend on the exact implementation, but assuming ten to twenty stacked metallic layers we estimate that the total thickness of the lines will be below \SI{5}{\micro \meter}. Polysilicon at temperatures close to zero Kelvin can have a thermal conductivity $\kappa$ =  \SI{100}{\watt / \meter / \kelvin}, and sufficient cooling power will be thus available at temperatures above $\approx$ 0.1K. Silicon metal-oxide-semiconductor (MOS) spin qubits can have a significant advantage for qubit operation at higher temperature, due to large energy scales of excited states and measured valley splittings, exceeding 10K \cite{Yang2013}. Further reductions in the required cooling power can be made by reducing the operation voltage, which is foreseeable at cryogenic temperatures, but possibly also by operating the transistors as single-electron-transistors \cite{Chen1996}, thereby significantly lowering the switching voltage. 

\section{Discussion}
The architecture shown here demonstrates that an array of single electron spins confined to quantum dots in isotopically purified silicon can be controlled using a scalable number of control lines. We have shown that the often argued compatibility of silicon spin qubits with standard CMOS technology is non-trivial. However, in the case of quantum dot qubits, the fabrication can be made consistent with standard CMOS technology and be scaled up to contain thousands to even millions of qubits. Provided that the down-scaling of CMOS transistors continues as anticipated, the control and measurement circuitry described can be integrated with qubits of a size that have already been experimentally demonstrated \cite{Veldhorst2014, Kawakami2014, Veldhorst2015}. The combination of ESR control, exchange coupling and dispersive readout of this design enables surface code operations to be performed using this platform. A key advantage is the possibility of global qubit control, so that many qubits can be addressed within the qubit coherence time. 

The proposed architecture is based on the current experimental status of silicon qubits and requires multiple transistors per qubit, significantly challenging CMOS manufacturing capabilities. Advancements in device uniformity and reproducibility could lower the number of required transistors. For example, with more uniform qubits the tuning circuitry and associated floating gates might not be needed. Additionally, operating at low magnetic fields will result in uniform qubit frequencies, avoiding the need for $g$-factor tuning. This limits functionality, since single-qubit gates can then be applied only globally, but universal computing is still possible using the local two-qubit gates. We anticipate that 2D arrays with such limited functionality can be realized in the near future, and will aid in the development of the universal quantum processor as presented here.  

The architecture for control and operation presented here is highly generic and can be implemented in a number of platforms, including spin qubits based on either Si/SiO$_2$ or Si/SiGe heterostructures, and various modes of operation such as single spin qubits \cite{Kawakami2014, Veldhorst2014}, singlet-triplet qubits \cite{Maune2012}, exchange-only \cite{Medford2013} or hybrid qubits \cite{Kim2014}. The system we considered here requires only local exchange interaction, but the architecture could also be incorporated in larger architectures that include long-range qubit coupling \cite{Petersson2012, Frey2012, Trifunovic2012,Braakman2013}, for example to interconnect quantum structures as presented here. While we consider the fabrication on a single SOI wafer, a more advanced and complex fabrication process could include multiple stacked layers to allow for more complex classical electronics per qubit, or for a separate control circuit that is purely dedicated for calibration and stability. A more sophisticated design could also include frequency multiplexing along a row, allowing global readout.  While the full fabrication and operation of our architecture is a formidable task, we believe that the detailed description together with the key requirements identified here pave the way towards an era of large-scale quantum computation; using the same silicon chip technology that has defined our current information age.

\section{Acknowledgements}
We thank Lieven Vandersypen for enlightening discussions. The authors acknowledge support from the Australian Research Council (CE11E0001017), the US Army Research Office (W911NF-13-1-0024) and the Commonwealth Bank of Australia.

\section{Author contributions}
M. V. and C.H.Y. designed the floating gates addressed via vertical transistors on top of the qubit plane. H.G.J.E. and M.V. made the design compatible with universal quantum computation. A.S.D. initiated and supervised the project. All authors provided input in writing the manuscript. 

The authors declare no competing financial interests.

Correspondence should be addressed to:

M.V. (M.Veldhorst@tudelft.nl) or

A.S.D.(A.Dzurak@unsw.edu.au).

\clearpage
\begin{widetext}

\begin{center}
\textrm{\textbf{Supplementary Information}}
\textrm{\textbf{CMOS archictecture for a 2D array of spin qubits}}
\end{center}

\textbf{Section 1. Functionality and qubit size} \\
The tremendous improvements in CMOS technology have resulted in feature sizes that are well below the minimal requirements for quantum dot definition. However, we envision that the small acceptable tolerance levels of qubits will require a certain amount of control lines for tunability. In a dense 2D array, these set of requirements will then determine the minimum qubit size for an extendable structure. Our proposal as described in the main text assumes a single floating gate for quantum dot definition and a single floating gate for qubit coupling between each qubit. These two gates are controlled by a circuit and are then extendable. The electrical circuit, as shown in Fig. 1b in the main text, includes six transistors that connect the data lines via the word lines and bit lines to the floating gates. For simplicity we have shown only one $J$-gate control structure, whereas an extendable structure contains two. The complete physical circuit corresponding to such an extendable element is shown in Supplementary Figure \ref{fig:s1} and has a footprint of 80 $\lambda^2$. Here, we have assumed that each line or line separation takes on a single linewidth $\lambda$. A feature size of 7 nm would then equate to a quantum dot size (including half the barrier area that separates the qubits) of $\approx 63 $nm$ \times 63$ nm. Small variations can be expected depending on the design considerations, but these quantum dots size are consistent with experimental realizations of quantum dot qubits \cite{Veldhorst2014supp, Kawakami2014supp}. \\ 

\begin{figure*} [h]
\includegraphics[width=0.9\textwidth]{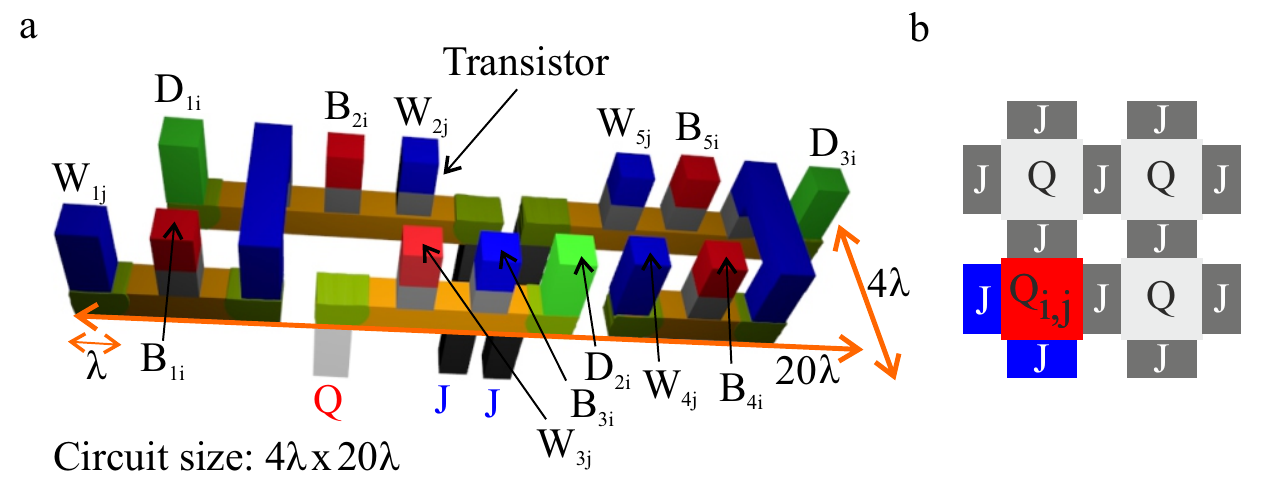}
\caption{\textbf{Control element, transistor circuit, a}. Control element for a single qubit and two $J$-gates (electrical circuit is depicted in the main text, Fig. 1b). The grey elements correspond to the transistor switches allowing to activate a line. The scale $\lambda$ is the features size, which is taken constant for each metal or dielectric layer. \textbf{b} Corresponding qubit and associated $J$ gates. 
\label{fig:s1}}
\end{figure*}

\textbf{Section 2. Matching planes (from qubit to address line)} \\
In drawing the physical qubit structure we have assumed a single linewidth, for both gate definition and gate separation. This allows us to define a single parameter $\lambda$, set by the feature size of the fabrication platform. While a 2D qubit plane takes on a square shape due to square (or circular) size of qubits, we found that this is generally not the case for the most optimal classical control layers. These different aspect ratios of the planes are matched using the vias, and can take on the same shape after expanding the qubits to a larger number, as described below.

We start with the basic control structure, which connects to a qubit and two $J$-gates, see Supplementary Figure \ref{fig:s2}. The aspect ratio of the control structure is $4 \lambda \times 20 \lambda$. In order to match with a square qubit, we extend the control structure to a set of 20 $\times$ 9, and the resulting structure is shown in Supplementary Figure \ref{fig:s3}. This control structure addresses a qubit array 20 $\times$ 4, which has the same footprint. However, in order to match the surface code protocol shown in the main text, Fig. 3, we again have to extend the structure to hold 54 $\times$ 9 classical control structures for $24 \times 20$ qubits (note the presence of 6 redundant classical control structures that appear after matching aspect ratio). The resulting, completely extandable, structure is shown in Supplementary Figure \ref{fig:s4}.

\begin{figure*}
\includegraphics[width=0.6\textwidth]{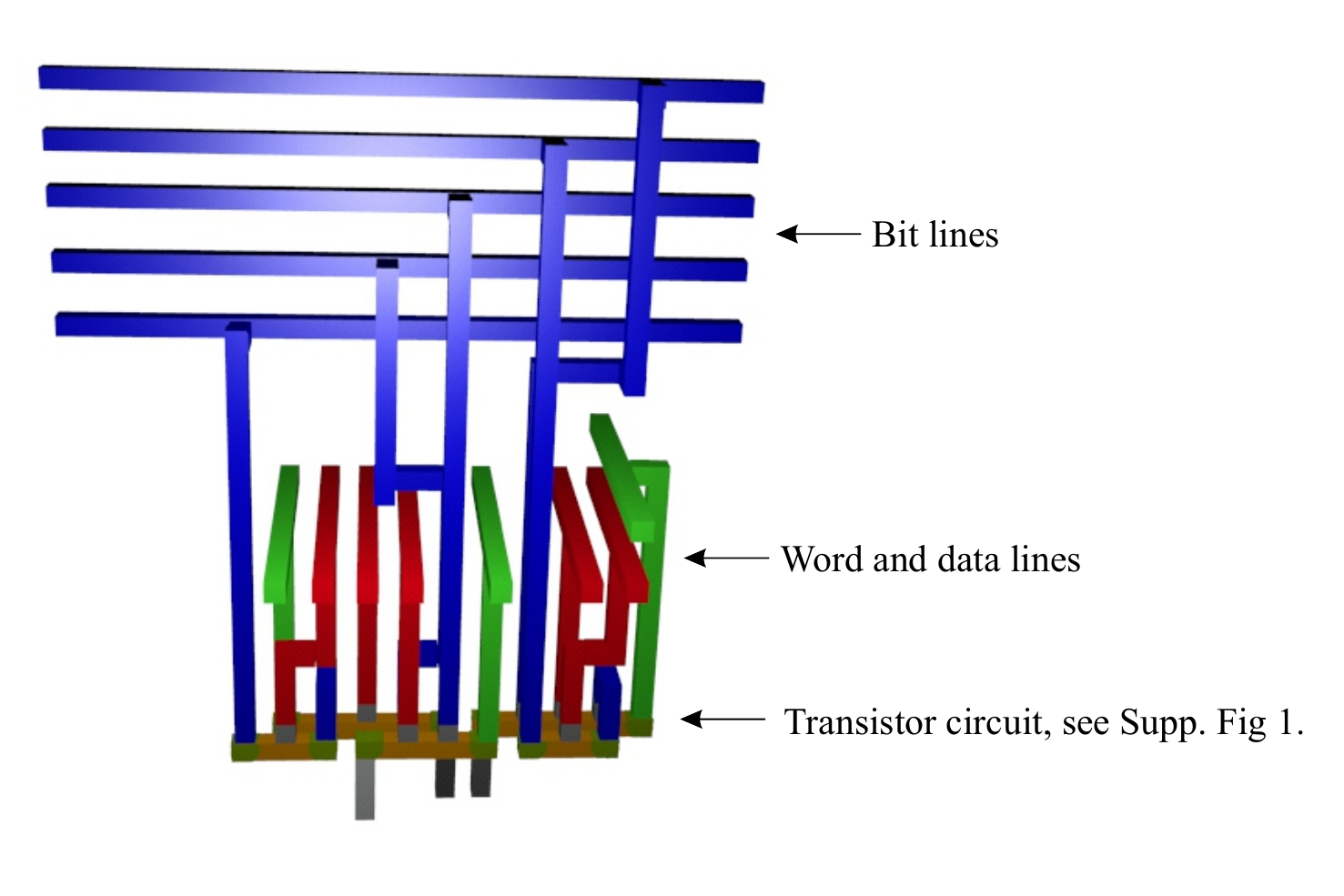}
\caption{\textbf{Control element, word and bit lines}. Control element for a single qubit (zoomed out version of Supplementary Figure \ref{fig:s1}). In a 2D quantum dot array with nearest neighbour coupling, the basic elementary scalable structure is one qubit, and two $J$-gates. 
\label{fig:s2}}
\end{figure*}

\begin{figure*}
\includegraphics[width=0.6\textwidth]{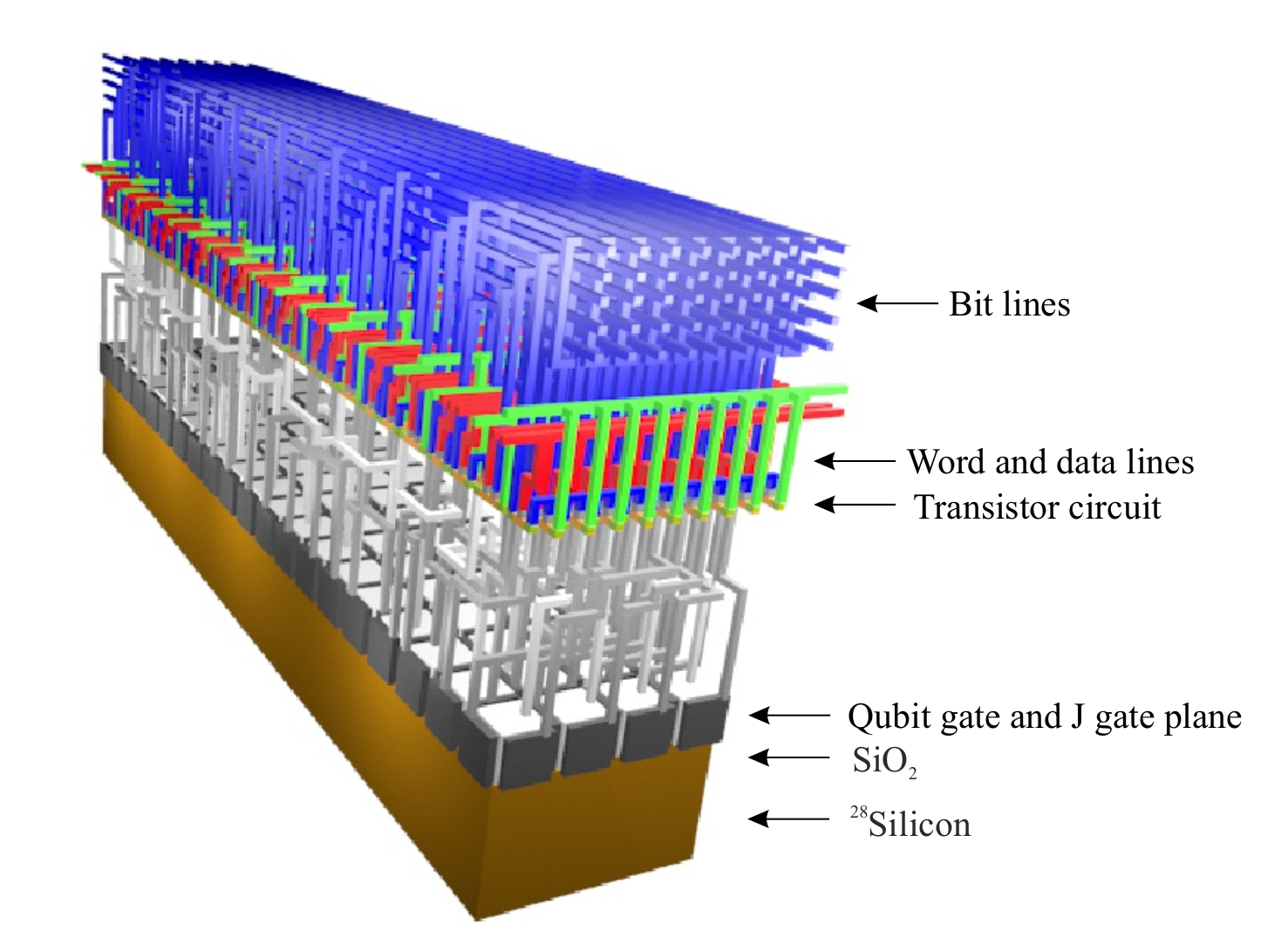}
\caption{\textbf{80 Qubit array}. The control element for a single qubit and two $J$-gates are extended to a $4 \times 20$ qubit array, in order to match the difference in aspect ratios between the qubits and control structures.
\label{fig:s3}}
\end{figure*}

\begin{figure*}
\includegraphics[width=1\textwidth]{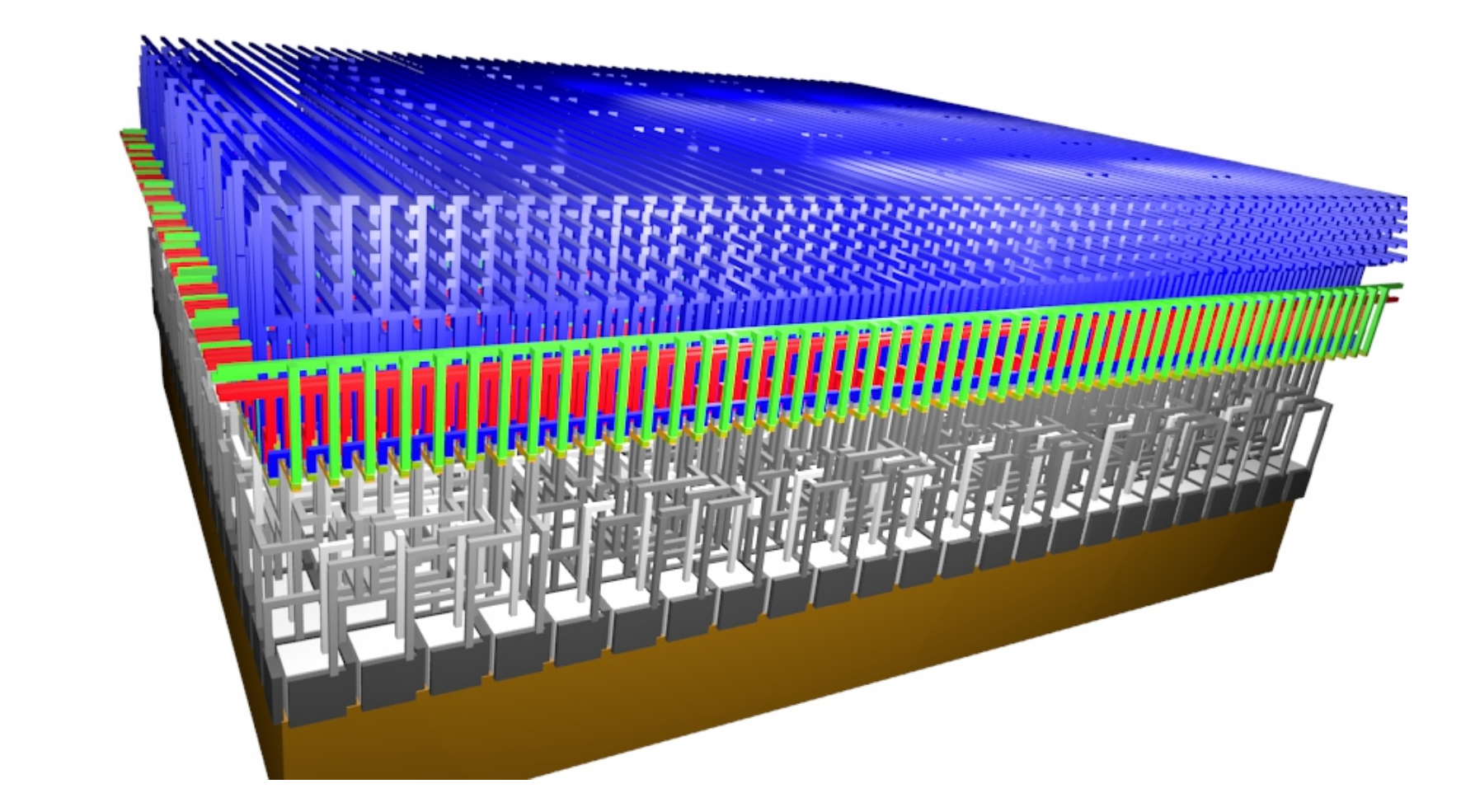}
\caption{\textbf{480 Qubit array}. In order to match the qubit array and the control structure with the surface code sequence operated via the associated Data lines, Word lines and Bit lines, the structure is extended to a $24 \times 20$ qubit array, which can be divided into an integer number of 6 qubits (which is the extendable structure for one surface code cycle, as discussed in the main text, Figure 3).
\label{fig:s4}}
\end{figure*}

\clearpage

\textbf{Section 3. Qubit operation} \\
In this section we discuss the qubit operation in more detail. \\ 

\textit{Qubit initialization and readout} \\
There are currently two popular methods for qubit readout based on spin-to-charge-conversion within the spin qubit community: readout based on the Zeeman energy (using a reservoir) and readout based on the singlet-triplet energy (via Pauli spin blockade). While both approaches can be made compatible with our control circuit, readout based on Pauli spin blockade offers the advantage of a larger energy scale (higher readout fidelity), no necessity for a large electron reservoir, and the qubit resonance frequency can be independently optimized (e.g. operation can be at low frequency).
Two electron spins residing in adjacent dots can be coupled by turning on the $J$-gate. In an adiabatic experiment, single spin states can then be converted onto the singlet-triplet axis. The triplet states have charge occupancy state (1,1), whereas the singlet states are in the (0,2) state. The resulting difference in capacitive coupling to the floating gate can be used for dispersive readout, i.e. the reflected power in an RF-setup will depend on the qubit spin state. Adiabatic separation of the two electrons initializes then the qubits for the following cycle. \\

\textit{Single-qubit logic operations} \\
We assume that the complete 2D-plane is positioned inside a cavity. In order to perform qubit operations, e.g. surface code operation, six individual qubit resonance frequencies are needed to individually control the qubit subsets (the $z$-axis qubits $Z1$ and $Z2$, the data-qubits $D1$ and $D2$, and the $x$-axis qubits $X1$ and $X2$) as shown in Fig. 3 of the main text. These operations are controlled globally via the cavity. Individual qubit tuning is controlled electrically via $g$-factor control \cite{Veldhorst2014supp}. This tuning will allow to calibrate the qubits into the required subsets, but also to actively (de)select qubit to create the 'holes', essential in surface code operation \cite{Fowler2012supp}. \\

\textit{Two-qubit logic operations} \\
Two-qubit operations are achieved via electrically controlling the tunnel coupling and/or detuning energy, experimentally realized in \cite{Veldhorst2015supp}. By turning the interaction on, the qubits will acquire a time-integrated phase dependent on the spin state of the coupled qubit. This allows to create either a SWAP or CPHASE operation, set by the interaction strength and the respective qubit resonance frequency difference. These two-qubit gates allows then to perform the surface code cycle, as shown in the main text, Fig. 3b. \\

\textit{Surface code with quantum dot spin qubits} \\
The general surface code cycle \cite{Fowler2012supp} is shown in Supplementary Figure \ref{fig:s5}a, which contains a sequence of CNOT operations together with single qubit Hadamards, readout and initialization steps. For the spin qubit approach, we realize these CNOT operations via a combination of CPHASE and two single qubit pulses, shown in Supplementary Figure \ref{fig:s5}b. For the readout phase, we implement two 'reference' qubits, which are another two quantum dots. The measurement qubits can then be readout via dispersive charge detection and the resulting sequence is shown in Supplementary Figure \ref{fig:s5}c. Here, we have included a SWAP operation, such that initialization can be adiabatically achieved using the tuned $g$-factor difference (see section 3 qubit initialization and readout for details).

\begin{figure*}
\includegraphics[width=0.9\textwidth]{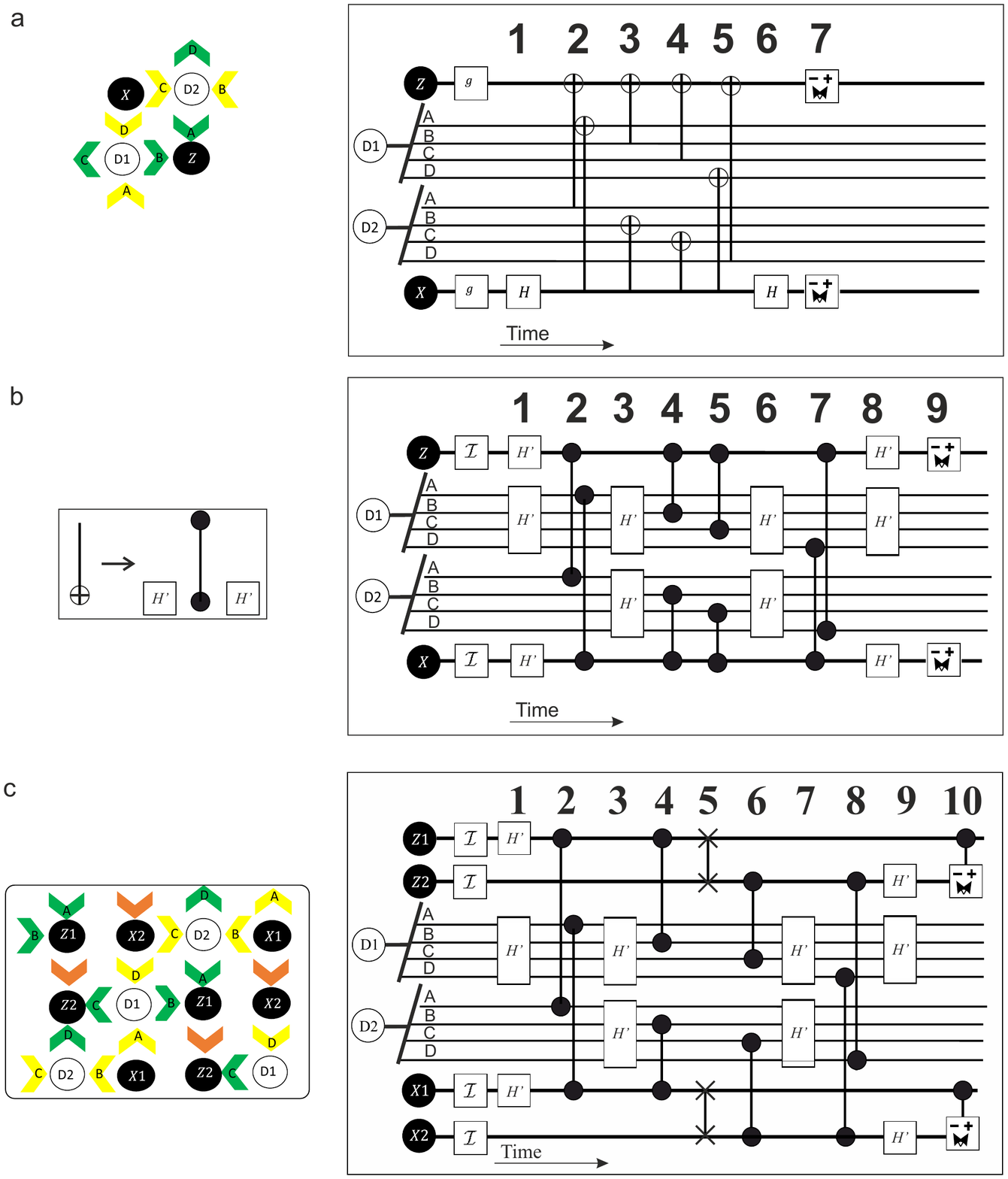}
\caption{\textbf{Surface code operation. a} General surface code operation.  \textbf{b} Surface code cycle after decomposing the CNOT into CPHASE and Hadamard operations. CPHASE operations with quantum dot qubits usually result in additional $\hat{z}$ rotations, which can be corrected using single qubit gates. Here this is included in the Hadamard, resulting in a Hadamard-like operation. \textbf{c} Surface code on a 2D quantum dot array.
\label{fig:s5}}
\end{figure*}

\end{widetext}

\end{document}